\documentclass[preprint,showpacs,preprintnumbers,amsmath,amssymb,prl]{revtex4}

\usepackage{graphicx}
\usepackage{dcolumn}
\usepackage{bm}

\begin{document}

\preprint{APS/123-QED}

\title{The effect of inhomogenities on single molecule imaging by
hard XFEL pulses}

\author{Zolt\'{a}n Jurek}
\author{Gyula Faigel}
\affiliation{Research Institute for Solid State Physics and Optics\\
             H 1525 Budapest, POB 49, Hungary}

\date{\today}

\begin{abstract}
We study the local distortion of the atomic structure in small
biological samples illuminated by x-ray free electron laser (XFEL)
pulses. We concentrate on the effect of inhomogenities: 
heavy atoms in a light matrix and 
non-homogeneous spatial distribution of atoms. In biological systems
we find both. Using molecular-dynamics type modeling it is shown
that the local distortions about heavy atoms are larger than the
average distortion in the light matrix. Further it is also shown
that the large spatial density fluctuations also significantly alter
the time evolution of atomic displacements as compared to samples
with uniform density. This fact has serious consequences on single
particle imaging. This is discussed and the possibility
of a correction is envisaged.  
\end{abstract}

                              

\maketitle

\section{\label{sec:intro}Introduction}

With the introduction of XFEL sources new possibilities open for
structural studies. This is clearly illustrated by the several works
done on the already existing soft XFEL-s
\cite{Chapman2007,Hajdu2008,Bogan2008}. In the near future similar
sources in the hard x-ray regime will also be operational.
According to the work of Neutze et al \cite{Neutze2000} the very
short and intense x-ray pulses of XFEL-s might allow atomic
resolution imaging of single non-periodic objects. Their idea is
that one collects the elastically scattered photons before the
deterioration of the sample is appreciable. Many models have been
worked out for the description of the behavior of the sample in the
XFEL pulse \cite{JurekEPJD2004,HauRiege2004,Bergh2004}. Their
conclusion is that at $\sim$10~keV the $\sim10^{13}$ photons have
to be concentrated in an approximately 10~fs time window for
successful atomic resolution imaging. Sources presently under
construction do not meet simultaneously all the requirements.
Therefore various ideas are introduced to relax the above
conditions. One of them is the application of a thin sacrificial
tamper layer about the sample
\cite{HauRiege2004,HauRiege2007,JurekEPJD2008}. In this case the
sample deteriorates with a slower rate, which may allow longer times
for imaging. However, this layer adds to the background, which might
hinder structure solution \cite{JurekEPJD2008,Bortel2007}. Other
suggestions concentrate to the evaluation process
\cite{JurekEPJD2008,Russel2008}: in one of them \cite{JurekEPJD2008}
the possibility to correct the charge loss during imaging is pointed
out, and in the other a new method for the construction of the 3D
reciprocal space dataset is introduced \cite{Russel2008}.  To
understand the usefulness and limits of the various approaches one
should start from as realistic description of the damage process as
possible. So far most of the modeling was done for spatially
homogeneous samples. The reason is that this assumption makes the
modeling simpler, more tractable. This allowed the application of
continuum models, with reduced dimensionality (1D)
\cite{HauRiege2004,HauRiege2007}. Even more sophisticated models
like Molecular Dynamics type models (MD) were applied for
homogeneous systems, mostly because of comparability with the
results of continuum models \cite{JurekEPJD2004,JurekEPJD2008}.
However, MD modeling allows the description of more realistic
non-homogeneous systems like biological molecules. Further, it has
an other advantage; it directly gives atomic coordinates, which are
necessary for the calculation of the diffraction pattern.  In this
paper we concentrate on the effect of inhomogenities on sample
dynamics. Inhomogenity is inherent to biological systems. In these
systems we find short and long near neighbor distances leading to
spatially non-homogeneous arrangement of atoms. Further, in many
cases we also find heavy elements in the light matrix. A typical
example is iron, which is contained by various building blocks of
living organisms. However, many other heavy atoms (Co, Ni, I etc.)
can also be found in biological systems.  In the present work we
show that the local distortion about heavy atoms is much larger than
the average distortion in a biological sample. It is also
shown that the large spatial density fluctuations significantly
alter the time evolution of the atomic displacements. This fact was
not recognized until now. We discuss our results from the point of
view of structural imaging. We use simple C and C--Fe model systems
to separately illustrate the above points, and we also show the
deterioration of a real biological system, the myoglobin, in which
the two effects appear at the same time.  Based on our
model-calculations it might be possible to correct for the above
distorting effects.

\section{\label{sec:model}Model parameters and model-samples}

In the proposed experiment, individual identical molecules are going
to be exposed to XFEL pulses one-by-one in random, unknown
orientation. Estimates for the number of elastically scattered
photons show, that single diffraction patterns will be very noisy.
Therefore the compilation of a 3D diffraction pattern, which is
necessary to solve the structure, is not possible directly from the
individual 2D images. Many pictures have to be collected and added
to improve statistics. This means that the minimum requirement of
photon counting statistics is determined by the ability to classify
the diffraction patterns according to their orientations. The
requirements set by the classification might be relaxed according to
a new approach, which uses the collected data set as a whole
\cite{Russel2008}.  However, the applicability of this approach is
not proved for a realistic large data set, therefore we use the more
conservative estimate given by the classification
\cite{Bortel2007,Bortel2009}. 
\begin{figure}
\includegraphics[scale=0.21]{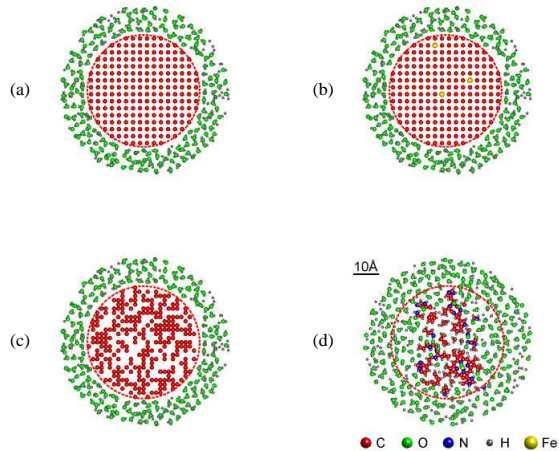}
\caption{\label{fig:F1}Starting structures used in the
model-calculations. For clarity we show four atomic layers thick
cross sections of the samples: homogeneous 40~\AA\ diameter pure
Carbon sample (a); the same sample as in (a) but 3 carbon atoms
replaced by Fe (b); spatially non-homogeneous sample (c), and
myoglobin (d). All samples are surrounded by a water layer resulting
in spherical droplets with 60~\AA\ total diameters. The sample parts
are encircled by 40~\AA\ diameter spheres. The color convention
given in the figure is used through the paper.}
\end{figure}
Based on \cite{Bortel2007,Bortel2009} we are considering XFEL pulses
with energy of 12~keV and with fluence $10^{13}$~ph/pulse. This beam
is focused to a 100~nm diameter spot. As pulse length we use 10~fs
with flat top shape, which is shorter than the pulse length of
XFEL-s under construction but it is realistically reachable. The
reason of using this short pulse is that even under this short time,
atoms might move appreciable distances.  In our MD model the motion
of all particles are followed by solving the non-relativistic
equations of motion. Coulomb forces are explicitly included, and the
various quantum-processes are taken into account through their cross
sections as stochastic processes. The detailed description of this
model is given in \cite{JurekEPJD2004}.  As a reference model-system
we use a 40~\AA\ diameter carbon cluster with homogeneously
distributed C-s, surrounded by a 10~\AA\ thick water tamper layer
[sample (A)] (Fig.~\ref{fig:F1}.a.).  Atoms are placed on a 2.5~\AA\
grid, which corresponds to the typical density of biological
systems, 1.35~g/cm$^3$.  In order to illustrate the effect of heavy
atoms, the same 40~\AA\ diameter carbon cluster is used with three
carbons exchanged to Fe [sample (B)] (Fig.~\ref{fig:F1}.b.). The
effect of spatial inhomogenities is studied on a pure carbon system
with similar sizes and average density as in the previous cases, but
in this sample the density is not spatially homogeneous [sample
(C)]. The sample is built in the following way: first C atoms are
placed on a dense regular grid $a=1.5$~\AA. This leads to
6.25~g/cm$^3$ density.  To reach the 1.35~g/cm$^3$, randomly chosen
atoms are removed (Fig.~\ref{fig:F1}.c). At last the behaviour of a
real biological molecule, the myoglobin is modeled [sample (D)]
(Fig.~\ref{fig:F1}.d ).

\section{\label{sec:results}Results}

\begin{figure}
\includegraphics[scale=0.21]{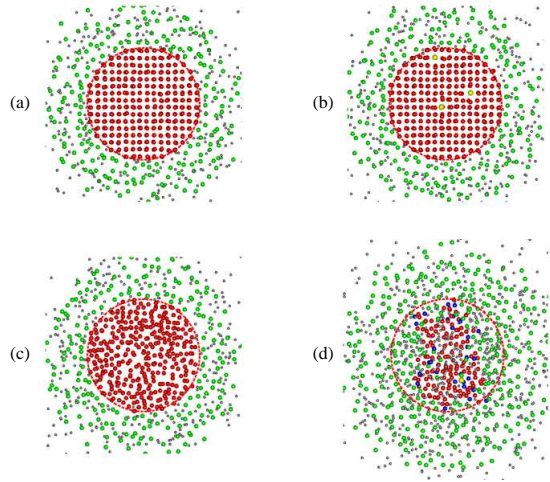}
\caption{\label{fig:F2} The structure of C (a), Fe--C  (b),
non-homogeneous C (c) and myoglobin (d) at the end of the pulse.}
\end{figure}

\begin{figure}
\includegraphics[scale=0.19]{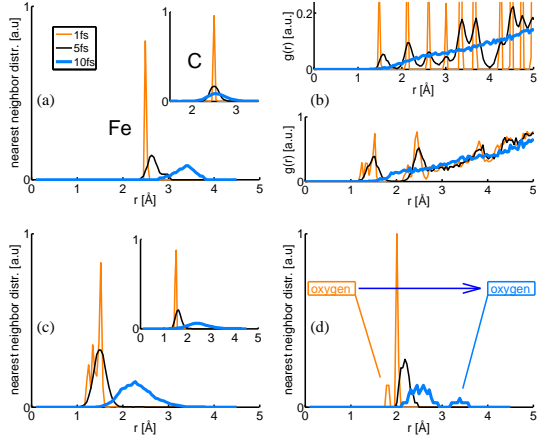}
\caption{\label{fig:F3} Time evolution of the first neighbor
distributions and pair correlation functions. (a) first neighbor
distributions around the Fe and C atoms (inset) in sample (B). (b)
pair correlation functions of sample (C) (upper panel) and (D) (lower
panel). (c) First neighbor distributions around the C atoms in
sample (D) and (C) (inset). (d) First neighbor distribution around
the Fe in sample (D).} 
\end{figure}

\begin{figure*}
\includegraphics[scale=0.5]{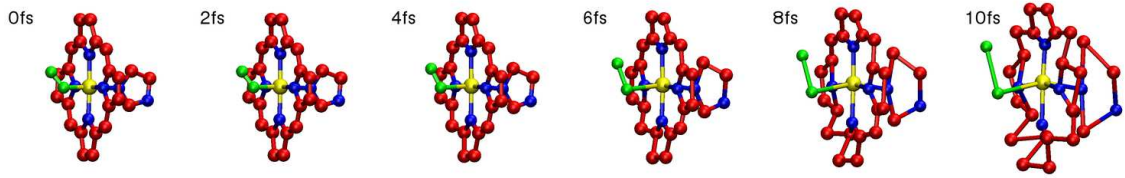}
\caption{\label{fig:F4} The 4.5\ \AA\  radius environment of the Fe
atom in the myoglobin at 0, 2, 4, 6, 8 and 10~fs.} 
\end{figure*}

First we discuss the effect of heavy atoms. On Fig.~\ref{fig:F2}.a
and b the atomic positions in the C and C--Fe systems are shown at
the end of the pulse. 
For easy comparison not the whole samples, only their cross sections
are depicted. Further, we do not show the electrons for clarity. In
the case of the pure carbon system [sample (A)] the changes in the
structure are hardly recognizable. However, in the water layer there
are significant distortions. The reason of the different behavior of
the sample and of the water parts were discussed in previous works
\cite{HauRiege2007,JurekEPJD2008}. Here we briefly describe the
processes leading to this particular behaviour. This discussion also
helps to understand the reasoning behind the effects of heavy atoms
and spatial inhomogenities. When the probe beam hits the sample,
high energy photoelectrons are produced. These leave the sample
resulting in highly excited ions and a positively charged particle.
Auger relaxation and electron-ion collisions lead to a large number
of free electrons. These non-bonded electrons rapidly rearrange in
such a way, that a neutral core -- composed of ions and electrons --
develops. In the core there is an effective Debye shielding, which
slows down the deterioration of the sample. Around the core we find
a positively charged layer, in which the Coulomb forces are not
shielded resulting in the fast explosion of this layer. In order to
avoid large structural distortion of the sample the thickness of the
water layer has to be chosen to be equal to the thickness of the
positively charged shell. In our case 10~\AA\ was chosen as the
minimum necessary thickness. Until this point our results agree with
those of the continuum models \cite{HauRiege2004,HauRiege2007}. Now
we turn to the Fe--C sample. Inspecting Fig.~\ref{fig:F2}.b. it is
clear, that a significantly larger distortion develops around the Fe
atoms than around the matrix C atoms. To make this difference more
quantitative we plotted the first nieghbour distributions about Fe
and C atoms on Fig.~\ref{fig:F3}.a. 
In the case of Fe centers the distortion is large; the center of the
distribution shifts about 1~\AA\ as compared to the original atomic
positions. Further, the width of the distribution increases to about
1~\AA. An increase of the width is expected because the random
nature of various ionization processes lead to different momentum
transfer to the ions resulting in different displacements.
For C atoms, the center of the distribution shifts only around
0.1~\AA, small compared to the Fe case. The increase of the width of
the distribution is about the same as for Fe. We can understand the
different behaviour of Fe environment by analyzing the physical
processes. The differences come from two sources: (i) the ionization
dynamics of the Fe is much faster than that of C, therefore a high
ionization state of the Fe atoms develops in a short time; (ii) the
Debye shielding does not work effectively for local inhomogenities.
The reason is that the electrons shield the average charge of the
system, which is determined by the C ions. The much higher positive
charge of the Fe ions is not compensated by the collective motion of
electrons.
These two factors lead to a large Coulomb repulsion about the Fe
atoms, which pushes out the neighboring C-s.  Next we describe the
effect of non-homogeneous spatial atomic arrangement.  On
Fig.~\ref{fig:F2}.c. sample (C) is shown at the end of the pulse.
We can identify large structural changes. One can follow this
quantitatively by analyzing the time dependence of the pair
correlation function (Fig.~\ref{fig:F3}.b.,upper panel) and the
nearest neighbour distribution (Fig.~\ref{fig:F3}.c. inset). We can divide
the nearest neighbours into three classes: 1.5~\AA, 2.1~\AA,
2.55~\AA. Following these separately, we see a large increase of the
peak position of the first class, a much smaller change in the
second class and almost nothing in the third class. At the end of
the pulse an almost homogeneous system develops. The explanation of
this behaviour is the following: the shielding by the free electrons
is determined by the average density, which corresponds to 2.5~\AA\
first neighbour distance. In this spatial range the Debye shielding
is effective. However, for shorter distances the positive charge of
the C ions are not compensated, which leads to large Coulomb forces
between close ions. The second factor, which determines the motion
of ions is the available space. In the inhomogeneous system the ions
tend to move toward those places, where the local density is small
(see Fig.~\ref{fig:F1}.c. and Fig.~\ref{fig:F2}.c).  Starting from
the above calculations we can understand the behaviour of the
myoglobin in the XFEL pulse. In Fig.~\ref{fig:F2}.d. the molecule is
shown at the end of the pulse. Its deterioration is severe, it
resembles to that of sample (C).
Plotting the time evolution of the pair correlation functions
(Fig.~\ref{fig:F3}.b.,lower panel) we see a similar trend than for
the model system, with the exception, that one cannot separate three
distinct first neighbour distances. This is natural, since in a real
molecule the bonding distances change almost continuously according
to the various types of bonds. We can obtain even more details by
separately studying the C and Fe environments (Fig.~\ref{fig:F3}.c
and d.).
While the time evolution of the C environment is very close to that
of the reference system (Fig.~\ref{fig:F3}.c.) the Fe environment
shows a slightly different behaviour (Fig.~\ref{fig:F3}.a. and d.).
The reason is that in the case of myoglobin the Fe first neighbour
distances are smaller than it was in the model system. Therefore the
heavy atom effect is combined with the effect of non-homogeneous
spatial arrangement.  As a result, some elements in the Fe
environment move even more than in the model system, and some move
less, depending on the interplay between the geometric arrangement
and the extra charge on the Fe ion. To illustrate this, we plotted
the environment of the Fe ion as a function of time
(Fig.~\ref{fig:F4}). 
The figure clearly shows that the oxygen atoms move the most, while
the nitrogen atoms the less. In the light of our calculations this
is not surprising since the Fe--O distance is the shortest leading
to a large Coulomb repulsion. Further, there is a large empty space
for the oxygen atoms to move to. In the case of the nitrogen atoms
the starting Fe--N distance is larger and there is no room around
the nitrogen atoms, so their neighbors slow down their motions.  At
last we point out one more interesting feature. In all cases the
volume and the shape of the samples do not change. To emphasize this
we put a constant diameter sphere around the samples
(Fig.~\ref{fig:F1}. and Fig.~\ref{fig:F2}.). Why is this fact
important? This clearly shows that a 1D homogeneous shell model
would predict a negligible distortion for all samples, since the
Debye shielding, which governs the damage in these models depends on
the average density but not on the detailed structure of the
samples.  However, as our calculations show this is very far from
reality. Of course our conclusions can be generalized to other
non-biological systems containing inhomogenities. This fact has
serious consequences on single particle imaging. Although the
overall shape and size of the samples may remain the same during an
XFEL pulse, their internal atomic structure can significantly
change. Therefore the reconstruction of the original structure
at the atomic level cannot be directly obtained, even slowing down
the sample destruction with the application of a tamper
layer. This layer slows down the explosion of the sample as a
whole, but the internal atomic structure may
drastically change.

\section{\label{sec:summary}Summary and conclusion}

The behaviour of small samples in highly focused XFEL
pulses was studied. We have examined the effect of inhomogenities on
the time evolution of the distortions of near neighbor environments.
It was found that around heavy atoms the local distortions
significantly increase as compared to the distortions about the
light matrix atoms. An even larger distortion develops around 
density fluctuations, where the light atoms are much closer than the
average distance corresponding to the density of the system. Both of
these effects are present in biological systems, which are one of
the target areas of research by XFEL sources. Until now most of the
damage modeling was done for homogeneous systems, and conclusions
about the feasibility of structural imaging experiments were drawn
from these calculations. Our study shows that atomic resolution
imaging is hindered by inhomogenities, and no reliable information
can be obtained from models using simplified homogeneous density.
Unfortunately this problem cannot be overcome by the application of
tamper layers, because local distortions show up even in the
presence of Debye shielding. However, knowing this type of behavior,
we can envisage the introduction of corrections in the evaluation
process. It is clear that we cannot avoid the distortions, but we
can use the fact that we know about it. We know that diffraction
data will result too large first neighbor distances about heavy atom
sites, and also about such sites where light atoms are very close to
each other. Using MD type modeling we can correct the structure
obtained directly from the diffraction data. This could be done in a
multi-step process: (i) roughly correct the structure obtained from
the diffraction data and use this as input parameters to the MD
modeling; (ii) calculate the time evolution of the corrected
structure and use this result to calculate the diffraction pattern;
(iii) compare this to the measured one and correct the starting
structure accordingly; (iv) start again MD modeling with this new
structure and continue until the calculated and measured patterns
are close enough. Using this type of approach, atomic resolution
imaging might be possible even in the presence of large local
distortions.
%

\begin{acknowledgments}

The work reported here was supported by OTKA 67866 and
NKFP1/0007/2005 grants.  

\end{acknowledgments}



\end{document}